\documentclass[12pt]{article}
\usepackage{amsmath,amsfonts,epsf}
\usepackage{amssymb}
\usepackage{graphicx}
\usepackage{grffile}
\input epsf

\def\be{\begin{equation}}
\def\ee{\end{equation}}
\def\bea{\begin{eqnarray}}
\def\eea{\end{eqnarray}}
\def\ie{\begin{equation}\begin{aligned}}
\def\fe{\end{aligned}\end{equation}}

\newcommand{\A}{{\alpha}}
\newcommand{\B}{{\beta}}
\newcommand{\C}{{\gamma}}

\renewcommand{\title}[1]{\vbox{\center\LARGE{#1}}\vspace{5mm}}
\renewcommand{\author}[1]{\vbox{\center#1}\vspace{5mm}}
\newcommand{\address}[1]{\vbox{\center\em#1}}
\newcommand{\email}[1]{\vbox{\center\tt#1}\vspace{5mm}}

\begin{document}
\begin{titlepage}
\begin{center}
\hfill \\
\hfill \\
\vskip 1cm

\title{Vector Models in the Singlet Sector at Finite Temperature}

\author{Stephen H. Shenker$^{a}$ and
Xi Yin$^{b}$}

\address{$^{a}$ Stanford Institute for Theoretical Physics and Department of Physics, Stanford University,  Stanford CA 94305 USA  \\  $^b$Center for the Fundamental Laws of Nature
Jefferson Physical Laboratory, Harvard University,
Cambridge, MA 02138 USA}

\email{$^a$sshenker@stanford.edu,
$^b$xiyin@fas.harvard.edu}

\end{center}

\abstract{ We study the thermal properties of the  $O(N)$ vector-like scalar theory in the singlet sector in $2+1$ dimensions.  This theory is conjectured to be the AdS/CFT dual of Vasiliev higher spin gravity.  We find that a large $N$ transition occurs but only at a very high temperature of order $\sqrt{N}$.   This corresponds to the bulk Planck energy. The transition signals a decrease in the number of degrees of freedom from that expected in the  simple higher spin gas, due to relations among the $O(N)$ bilinear invariants.}
\vfill

\end{titlepage}

\section{Introduction}
Substantial progress has been made in recent years in studying AdS/CFT dual pairs where the boundary field theory is a vector-like large $N$ system \cite{Klebanov:2002ja, Sezgin:2002rt, Petkou:2003zz, Sezgin:2003pt, Girardello:2002pp, Giombi:2009wh, Giombi:2010vg, Koch:2010cy, Douglas:2010rc, Giombi:2011ya, Gaberdiel:2010pz, Gaberdiel:2011wb, Gaberdiel:2011nt, Ahn:2011pv, Chang:2011mz, Papadodimas:2011pf}.  Here the boundary field theory can be analyzed by standard field theoretic techniques while the bulk theory is a gravitational theory with an infinite tower of higher spin fields.\footnote{A very  interesting recent proposal for a dS/CFT dual has been made in \cite{Anninos:2011ui}.} The higher spin bulk fields are dual to singlet bilinear operators in the field theory that can be written schematically as $\phi_i \partial \ldots \partial \phi_i$ \cite{Klebanov:2002ja}.
More precisely, a version of the conjecture in $2+1$  boundary dimensions relates Vasiliev's minimal bosonic higher spin gauge theory in $AdS_4$ to the CFT of $N$ free massless scalars $\phi_i$ restricted to the $O(N)$ singlet sector. The singlet constraint requires a bit of explanation. On local operators, it simply means projecting onto $O(N)$ invariant operators constructed out of $\phi_i$. Equivalently, one may couple the $N$ scalars to an $O(N)$ Chern-Simons gauge field at level $k$, and take the $k\to \infty$ limit in which the gauge dynamics decouple. This description has a well defined local Lagrangian. Note that the Chern-Simons level is not renormalized quantum mechanically, and such Chern-Simons-matter theories define CFTs at general values of $k$.\footnote{The question about dynamics at finite $\lambda = N/k$ will be discussed in \cite{CSfermion}.} This is a feature unique to three dimensions, and does not hold for instance in two dimensions, where the analogous $O(N)$ projection also requires twisted sector states by modular invariance.  A similar conjecture relates Vasiliev's non-minimal bosonic higher spin gauge theory in $AdS_4$ to the theory of $N$ complex massless scalars in the $U(N)$ invariant sector. Other versions of the conjecture involve critical $O(N)$ vector models or their $U(N)$ generalizations, which are interacting CFTs. The duals to the free CFT and the interacting fixed points differ by a choice of bulk boundary conditions on the bulk scalar \cite{Witten:2001ua, Klebanov:2002ja, Hartman:2006dy, Giombi:2011ya}. Nontrivial agreement between the three point functions of the bulk and boundary theories have been found \cite{Giombi:2009wh, Giombi:2010vg}.
  
One of the most important aspects of the AdS/CFT duality is the relationship between the  thermal behavior of the field theory and black holes in the bulk \cite{Witten:1998zw}.  For the field theory  on a sphere with dynamical fields in the adjoint representation there is typically a large $N$ phase transition at temperature $T \sim 1$ between a phase with entropy $S \sim 1$ and a high temperature phase with entropy $S \sim N^2$.  In the bulk the low temperature phase corresponds to an AdS gas of perturbative quanta.  The high temperature phase corresponds to a large AdS black hole and the entropy is that of the black hole with $G_N \sim 1/N^2$.  In the context of the higher spin gauge theory/vector model correspondence, a natural question to address is the thermal behavior of the boundary vector-like large $N$ system.  This note is a first step in that direction.    We focus on the $2+1D$  boundary theory conjectured to be dual to Vasiliev gravity \cite{Vasiliev:1992av, Vasiliev:1995dn, Vasiliev:1999ba, Vasiliev:2003ev, Sezgin:2002ru, Sezgin:2003pt}.  Related work on the matching of CFT entropy  in the $1+1D$ $W_N$ conformal field theory with that of the BTZ black holes in the $ 2+1D$  bulk  dual  has been discussed in \cite{Kraus:2011ds}.   

Our main findings are simple to state:  in this system there is a large $N$ thermal transition but it occurs at temperatures $ T\sim \sqrt{N}$, not $T \sim 1$.  In bulk units this corresponds to an energy of order Planck scale, not AdS scale. This transition reflects a decrease in the number of degrees of freedom compared to the bulk higher spin gas. This decrease is due to relations between products of bilinears which are treated as independent in the higher spin gas. This indicates the absence of  a thermodynamically stable large AdS-Schwarzschild black hole solution in this theory at $T \sim 1$. 

Thermodynamically subleading black holes should produce corrections to the dominant free energy of order $\exp(-1/G_N)$ or since $G_N \sim 1/N$ of order $\exp(-N)$.  But we find that in the case of the free scalar boundary theory at temperatures $T\sim 1$ the corrections to the free energy on the sphere are not only nonperturbative in $G_N$, but  of order  $ \exp(-N^{3/2})$, too small to be caused by a black hole. This indicates the absence of  (uncharged) AdS-Schwarzschild-like black hole solutions in Vasiliev's theory, at least with such boundary conditions.

In section 2 we discuss the thermal properties of the free $U(N)$ scalar gas at temperatures of order 1  using matrix techniques.  We then show that this answer agrees with the free bulk higher spin gas in the thermal AdS geometry.  In Section 3 we find a Gross-Witten phase transition at temperatures of order $\sqrt{N}$, corresponding to the bulk Planck energy.  In Section 4 we discuss the interacting, critical, boundary theory and show that this system behaves in the same way. In Section 5 we discuss finite $N$ relations between invariants and corrections. We then end with a brief discussion.

\section{Finite temperature system }
\subsection{Field theory partition function}
  We consider the Euclidean functional integral on $S^2 \times S^1$ of the theory with $N_f$ fundamental scalar multiplets scalar coupled to a $U(N)$ gauge theory\footnote{We work with $U(N)$ rather than $O(N)$ to simplify formulas.  Nothing of importance is affected.} with Chern Simons action at level $k \rightarrow \infty$.   $N_f$ is fixed as $N \rightarrow \infty$.   The necessary technology for this calculation has been developed in \cite{ Sundborg:1999ue, Aharony:2003sx}  in their benchmark study of finite temperature weakly coupled gauge theory on the sphere.   Their techniques have been applied to matter in the fundamental in \cite{Schnitzer:2006xz}. The integration over the $U(N)$ holonomy around the thermal $S^1$ serves to enforce the $U(N)$ singlet constraint.  There is no other gauge dynamics at $k \rightarrow \infty$.  Integrating out the scalar fields yields the following partition function for the $U(N)$ holonomy matrix, whose eigenvalues are denoted $\A_i$.
\ie \label{partfcn}
Z(N,\beta) &= \int [dU]_{U(N)} \exp\left\{ N_f \sum_{m=1}^\infty {1\over m} z_S(x^m) \left[ {\rm Tr}_\Box (U^m) + {\rm Tr}_{\bar\Box} (U^m) \right] \right\}
\\
&= {1\over N!} \int \prod_i d\A_i \exp\left[ \sum_{i<j} 2\ln |\sin({ \A_{i}- \A_{j} \over 2})| + 2 N_f \sum_{m=1}^\infty {1\over m} z_S(x^m) \sum_i \cos(m\A_i) \right]
\fe
where $ x = \exp(-1/T)$ and
\ie
z_S(x) = x^{{d\over 2}-1} {1+x\over (1-x)^{d-1}}
\fe
is the one letter partition function introduced in \cite{Aharony:2003sx}.   $d$ is the spacetime dimension of the boundary field theory.  For the theory on $S^2 \times S^1$, $d = 3$.

The higher spin bulk states are all $U(N)$ invariant bilinears of the scalars, and there are $N_f^2$ such bilinears.  So all thermal quantities given by the higher spin bulk gas will be exactly proportional to $N_f^2$.    As we will discuss below  relations  between invariants will be important at high enough temperatures.  Then the free energy will not have a precise $N_f^2$ dependence.  The change in this dependence will serve as an indicator of a large $N$ phase transition. 

To solve for the the free energy at large $N$ we determine the saddle point eigenvalue density $\rho(\A)$ \cite{Brezin:1977sv, Gross:1980he}.  Denoting the integrand in  (\ref{partfcn}) as $\exp(S[\rho])$ we can write
\ie \label{action}
S[\rho] = 
 N^2 \int d\A d\B \rho(\A) \rho(\B) \ln |\sin({\A-\B \over 2})| + 2 N_f  N \int d\A \rho(\A) \sum_{m=1}^\infty {1\over m} z_S(x^m) \cos(m\A).
\fe
We immediately see a difference from the case with adjoint matter that produced the ``Hagedorn/Deconfinement  transition" discussed in \cite{ Sundborg:1999ue, Aharony:2003sx}, as well as from the case with fundamental matter but with $N_f/N$ finite \cite{Schnitzer:2006xz}.  The measure factor for the eigenvalues (the Van der Monde determinant) is of order $N^2$.  Because there are only $N$ scalars the term in $S$ due to them is only of order $N_f N$.   So the saddle will not be substantially affected.  We can still compute the leading order 1  effect of the matter when $N$ is large  by computing the small shift in the saddle.  These results will be a special case of those in  \cite{Schnitzer:2006xz}.
Without the scalars this density is uniform $\rho(\A)= {1 \over 2 \pi}$.   Write the general saddle point as
\ie
\rho(\A) = {1\over 2\pi} + {N_f\over N} \tilde\rho(\A),
\fe 
Then $\tilde\rho(\A)$ solves the saddle point equation \cite{Gross:1980he}
\ie \label{spe}
{\bf P} \int d\B \tilde\rho(\B) \cot(\frac{\A-\B}{2}) - 2\sum_{m=1}^\infty z_S(x^m) \sin(m\A) = 0,
\fe
Observing that \cite{Gross:1980he}
\ie 
\cot((\alpha -\beta)/2) = 2 \sum_{n=1}^{\infty} (\sin (n \alpha) \cos(n \beta) - \cos(n \alpha) \sin(n\beta))
\fe
We can immediately find a solution for (\ref{spe})
\ie  \label{saddle}
{\tilde \rho}(\beta) = \sum_{m=1}^{\infty} z_S(x^m) \frac{1}{\pi}\cos(m \beta)
\fe

We can evaluate the $N_f$ dependent part of the free energy ${\cal F} = \log Z$ by computing the derivative
\ie\label{dfdn}
\frac{\partial {\cal F}}{\partial N_f} 
= 2N \int d\A[ \rho(\A) \sum_{m=1}^\infty
 {1\over m} 
 z_S(x^m) \cos(m\A) ]
\fe
Plugging in (\ref{saddle}) and noting the constant part of $\rho$ does not contribute we find
\ie
\frac{\partial {\cal F}}{\partial N_f} 
= 2N_f\sum_{m=1}^\infty[
 {1\over m} 
 z_S^2(x^m)]
\fe
Dropping an $N_F$ independent constant

\ie \label{freeenergy}
{\cal F}
= N_f^2\sum_{m=1}^\infty
 {1\over m} 
 z_S^2(x^m) =N_f^2\sum_{m =1}^{\infty}{1 \over m} {x^m (1+x^m)^2 \over (1-x^m)^4 }
\fe

As long as this calculation is valid the answer will always be exactly proportional to $N_f^2$.    (The bulk free higher spin gas corresponds to $N_f=1$.)

\subsection{Partition function of thermal higher spin gas}

We now directly  determine the bulk free higher spin gas on the thermal $AdS_4$ background to compare to the above calculation. One-particle states of a higher spin particle organize into an irreducible representation of the AdS isometry group $SO(3,2)$, labeled by the energy and spin of the lowest state. The free energy of the free higher spin gas is
\ie
{\cal F} =-  \sum_{s=0}^\infty \sum_{\ell=0}^\infty n(s,\ell) \ln (1-x^{s+1+\ell})
\fe
where $n(s,\ell)$ are the coefficients of the character of short representations of $SO(3,2)$ of $(\Delta, J) = (s+1,s)$, except the $s=0$ case where the representation labeled by $(1,0)$ is a long representation. Namely, for $s>0$,
\ie
\chi_{s+1,s}(q) = \sum_{\ell=0}^\infty n(s,\ell) q^{s+1+\ell} = G_{s+1,s} - G_{s+2,s-1},
\fe
where $G_{\Delta, s}$ is the character of a long representation,
\ie
G_{\Delta, s} = {(2s+1) x^\Delta\over (1-x)^3},
\fe
and in the special case $s=0$,
\ie
G_{1,0} = \sum_{\ell=0}^\infty n(0,\ell) q^{\ell+1}.
\fe
So we can write
\ie\label{freeHSgas}
{\cal F} = \sum_{m=1}^\infty {1\over m} \left[ G_{1,0}(x^m) + \sum_{s=1}^\infty \chi_{s+1,s}(x^m) \right]
= \sum_m {1 \over m} {x^m (1+x^m)^2 \over (1-x^m)^4 }.
\fe
This agrees with the boundary field theory calculation.

\section{Gross-Witten transition}

At first glance this calculation is valid for all temperatures because the matter term in (\ref{action}) is order $N$ while the measure is order $N^2$ and the saddle point is not changed by very much.  But this is not the case.   At very high temperatures there are so many scalar quanta excited that the terms can become comparable.    At very high temperatures compared to the AdS scale $(1-x)  \rightarrow 1/T$. So, setting $d=3$
\ie\label{zhit}
z_S(x^m) \approx 2(\frac{T}{m})^2.
\fe
The crucial point is that when $T \sim \sqrt{N}$,   $z_S \sim N$ and the matter term in (\ref{action}) is also order $N^2$.  So we are led to study high temperatures
$T = b \sqrt{N/N_f}$.  Then
\ie
z_S(x^m) \approx {1\over m^2} {2N\over N_f} b^2.
\fe
The saddle point equation becomes
\ie
 {\bf P} \int d\B \rho(\B) \cot(\frac{\A-\B}{2}) - 4b^2 \sum_{m=1}^\infty {\sin(m\A)\over m^2}  = 0.
\fe
The saddle point density $\rho_s$ is given by 
\ie
\rho_s(\alpha) = \frac{1}{2\pi} + \frac{2 b^2}{\pi} f(\alpha),
\fe
where 
\ie\label{falpha}
f(\alpha) = \sum_{m=1}^\infty \frac{\cos(m \A)}{m^2} = -\frac{\pi^2}{12} +\frac{(\A-\pi)^2}{4}.
\fe
Here we see the density changes appreciably when $b \sim 1$.  In fact  a Gross-Witten \cite{Gross:1980he} type transition occurs.  Above the transition the density is zero on a finite interval of $\A$.   The signal for this is $\rho_s(\alpha)$ vanishing somewhere.
This occurs first at $\alpha = \pi$ giving a critical temperature $T_c = b_c \sqrt{N/N_f}$, with
\ie
b_c = \frac{\sqrt{3}}{\pi}.
\fe
At larger $b$ the true eigenvalue density is zero over a finite range of $\A$.   At very high temperatures, $b \rightarrow \infty$,  all the $\A_i \sim 0$. Using ({\ref{dfdn}) we find
\ie\label{htl}
{\cal F} \approx 4\zeta(3)  N_f N T^2,
\fe
the unconstrained free field value, linear in $N_f$.    The endpoints of the eigenvalue region vary analytically with $b$ so ${\cal F} $ is analytic in $T$ above the transition.   In particular, ${\cal F}$ cannot be exactly proportional to $N_f^2$  in any region of  this high temperature phase.  This means it is not the higher spin gas phase.    

This phase transition is different than the one in standard large $N$ gauge theory.  There the system jumps from a phase with entropy order 1 to a phase with entropy order $N^2$.   
In the limit where the temperature is high but still of order $1$ compared to powers of $N$, we have from (\ref{freeenergy})
\ie \label{orderone}
{\cal F}(T) \approx 4\zeta(5) N_f^2  T^4.
\fe
So if we naively continue (\ref{orderone}) to higher temperatures we see that  at $T_c  = b_c \sqrt{ N \over N_f}$, the entropy is of order $N^2$ in both phases.  
At temperatures  $b \gg 1$ (\ref{orderone}) is order $b^4 N^2$, much larger than (\ref{htl}) which is of order $b^2 N^2$, suggesting that degrees of freedom are eliminated from the simple higher spin gas picture, rather than added.  We note that at the transition the entropy is of order $N^2$. 

The location of the transition $T \sim \sqrt{N}$ corresponds to Planckian energies in the bulk.  It is a surprise that the first thermally stable bulk configuration, the analog of a large AdS black hole, occurs at Planckian scales, not AdS scales.   This is true for singlet vector models in general dimensions.   In general $d$ (\ref{zhit}) becomes $z_S(x^m) \approx ({T \over m })^{(d-1)}  $.  So $T_c \sim N^{1 \over d-1}$.
Let $T_{ij}$ be the stress tensor of the CFT, which couples to the bulk graviton.   Normalize it so $\langle T T \rangle \sim 1$.   Then $\langle TTT \rangle \sim 1/\sqrt{N}$ in all dimensions.   This means Newton's constant $G_N \sim 1/N$ in all dimensions.  $G_N \sim m_P^{-(d-1)}$  (where the bulk dimension is $d+1$).  So $m_P \sim N^{1 \over d-1} $, the same order as $T_c$.

The $U(N)$ free fermion system works similarly. 
Here the matter part of $S$ is 
\ie \label{actionf}
2N_f\sum_{m=1}^\infty {(-1)^{m+1}\over m} z_F(x^m) \sum_i \cos(m\A),
\fe
where $z_F(x)$ is the fermion one letter partition function,
\ie
z_F(x) = \frac{2^{ {d  \over  2} +1}
x^{{d \over 2} - {1 \over 2}}}{
{(1-x)^{d-1}}}.
\fe
So for large $T$  (\ref{actionf}) becomes
\ie
V(\A)= N  2^{{d\over 2} + 2} T^{{d-1}}\sum_{m=1}^\infty \frac{(-1)^{m+1}\cos(m \A)}{m^d}.
\fe
Again, the Gross Witten transition happens at $b_c \sim 1$ for $T= bN^{1 \over d-1}$.  This means $T_c \sim m_P$ in general dimension.

A significant difference between fermions and bosons is the behavior of $V$ near $\A = \pm \pi$, the location of importance for the Gross Witten transition.  In particular for fermions in $d=3$ 
\ie
{d^2 V(\pi) \over d \A^2} 
\sim -\sum_{m=1}^\infty {1 \over m}
\fe
While for bosons it is
\ie
{d^2 V(\pi) \over d \A^2} \sim - \sum_{m=1}^\infty {(-1)^{m+1} \over m}
\fe
The divergent second derivative may affect the order of the Gross Witten transition but not its presence.

\section{Partition function of critical vector model in the singlet sector}

For the interacting,  critical $U(N)$ model, we need to replace the free scalar free energy (in the presence of Wilson lines)\footnote{The computation in this section generalizes that of \cite{Sachdev:1993pr}.}
\ie
&{\rm Tr} \ln (-\Delta_{S^2}(\A) + {1\over 4})
=  \sum_{i=1}^N \sum_{n=-\infty}^\infty \sum_{\ell=0}^\infty (2\ell+1) \ln \left[ (\ell+{1\over 2})^2 + ({2\pi n+ \A_i\over \beta} )^2  \right]
\fe
by
\ie\label{crif}
& {\rm Tr} \ln (-\Delta_{S^2}(\A) + {1\over 4} + \lambda) - 4\pi\beta N{\lambda\over g}
\\
&= \sum_{i=1}^N \sum_{n=-\infty}^\infty \sum_{\ell=0}^\infty (2\ell+1) \ln \left[ (\ell+{1\over 2})^2 + ({2\pi n+ \A_i\over \beta} )^2 + \lambda \right] - 4\pi\beta N{\lambda\over g}
\\
\fe
where the critical coupling $g$ is determined by
\ie
{1\over g} = \int {d^3p\over (2\pi)^3} {1\over p^2},
\fe
and the expectation value of the Lagrangian multiplier field $\lambda=\lambda(\A)$ is such that (\ref{crif}) is extremized,
\ie\label{sadlam}
 \sum_{i=1}^N \sum_{n=-\infty}^\infty \sum_{\ell=0}^\infty {2\ell+1 \over (\ell+{1\over 2})^2 + ({2\pi n+ \A_i\over \beta} )^2 + \lambda(\A) } = {4\pi\beta N\over g}.
\fe
In writing this saddle point equation, a UV regularization on both sides is understood.
In the matrix model language, going to the critical model amounts to replacing
\ie
z_S(x) = x^{1\over 2}{1+x\over (1-x)^2} = \sum_{\ell=0}^\infty (2\ell+1) x^{\ell+{1\over 2}}
\fe
by
\ie
z_{S,\lambda}(x) = \sum_{\ell=0}^\infty (2\ell+1) x^{\sqrt{(\ell+{1\over 2})^2+\lambda}}.
\fe
By a rewriting of the functional determinant with Wilson line, following \cite{Aharony:2003sx}, the matrix model for $U(N)$-invariant partition function becomes
\ie\label{zmcr}
Z(N,\beta) = \int d\lambda \int [dU]_{U(N)} \exp\left\{  \sum_{m=1}^\infty {1\over m} z_{S,\lambda}(x^m)
\left[ {\rm Tr}_\Box (U^m) +{\rm Tr}_{\overline \Box} (U^m) \right]  + \beta N \left[ {\cal F}_0(\lambda) + 4\pi{\lambda \over g}\right] \right\}
\fe
where 
\ie
{\cal F}_0(\lambda) = 4\pi \int {d^3p\over (2\pi)^3}\ln p^2 - \int {d\omega\over 2\pi} \sum_{\ell=0}^\infty (2\ell+1) \ln \left[ (\ell+{1\over 2})^2 + \omega^2 + \lambda \right].
\fe
The first term in ${\cal F}_0(\lambda)$ amounts to subtracting off a divergence that is present in flat spacetime.
We have
\ie \label{fzer}
{\cal F}_0(\lambda) + 4\pi{\lambda \over g} = \int {d\omega\over 2\pi} \left\{ 4\pi\int {d^2p\over (2\pi)^2}  \left( \ln p^2+{\lambda\over p^2}\right) - \sum_{\ell=0}^\infty (2\ell+1) \ln \left[ (\ell+{1\over 2})^2 + \omega^2 + \lambda \right]\right\}
\fe
which is a finite expression (after both the integral and the sum are regularized appropriately).

Now if we evaluate the potential at the uniform eigenvalue distribution, we obtain a nonzero piece proportional to $\beta N$, namely $\beta N$ times (\ref{fzer}) extremized with respect to $\lambda$. This contributes to a ground state energy proportional to $N$, and should be shifted away when comparing to the $AdS_4$ dual. After doing so, the conclusion remains that the free energy at temperature $T\ll \sqrt{N}$ is of order 1.

Let us examine the large $N$ limit of this matrix model in some more detail. In terms of the eigenvalue density $\rho(\A)$, the saddle point equation for $\lambda$ (\ref{sadlam}) takes the form
\ie
&\int d\A \,\rho(\A) \sum_{n=-\infty}^\infty \sum_{\ell=0}^\infty {2\ell+1\over (\ell+{1\over 2})^2 + ({2\pi n+\A\over \B})^2 + \lambda} 
\\
&= {\B\over 2}\int d\A \rho(\A) \sum_{\ell=0}^\infty {\ell+{1\over 2}\over \sqrt{(\ell+{1\over 2})^2+\lambda }} \left[ \coth\left({\B\over 2}\sqrt{(\ell+{1\over 2})^2+\lambda}+{i\A\over 2} \right)+\coth\left( {\B\over 2}\sqrt{(\ell+{1\over 2})^2+\lambda}-{i\A\over 2} \right) \right]
\\
& = 2\pi \B \int {d^2\vec p\over (2\pi)^2} {1\over |\vec p|},
\fe
where we have integrated out the momenta in Euclidean time direction on both sides.
The regularized version of this equation can be written as
\ie\label{sado}
& \int d\A \rho(\A) \sum_{\ell=0}^\infty \left\{ {\ell+{1\over 2}\over \sqrt{(\ell+{1\over 2})^2+\lambda }} {\coth\left( {\B\over 2}\sqrt{(\ell+{1\over 2})^2+\lambda}+{i\A\over 2} \right)+\coth\left( {\B\over 2}\sqrt{(\ell+{1\over 2})^2+\lambda}-{i\A\over 2} \right) \over 2} -1 \right\}
\\
& = 0.
\fe
The saddle point equation for $\rho(\A)$ is the one for the free theory with $z_S(x)$ replaced by $z_{S,\lambda}(x)$. The solution is given by
\ie\label{sadt}
&\rho(\A) = {1\over 2\pi} + {1\over N}\tilde\rho(\A),
\\
&\tilde\rho(\A) = \sum_{m=1}^\infty z_{S,\lambda}(x^m) {\cos(m\A)\over \pi}.
\fe
Now using the expansion
\ie
& {\coth\left( {\B\over 2}\sqrt{(\ell+{1\over 2})^2+\lambda}+{i\A\over 2} \right)+\coth\left( {\B\over 2}\sqrt{(\ell+{1\over 2})^2+\lambda}-{i\A\over 2} \right) \over 2} 
\\
&= 1+2\sum_{n=1}^\infty e^{-n\beta\sqrt{(\ell+{1\over 2})^2+\lambda}} \cos(n\A),
\fe
we can write (\ref{sado}) as
\ie\label{sadlll}
\sum_{\ell=0}^\infty \left[{\ell+{1\over 2}\over \sqrt{(\ell+{1\over 2})^2+\lambda}}-1\right]+{2\over N}\sum_{m=1}^\infty z_{S,\lambda}(x^m) \sum_{\ell=0}^\infty {\ell+{1\over 2}\over \sqrt{(\ell+{1\over 2})^2+\lambda}} x^{m\sqrt{(\ell+{1\over 2})^2+\lambda}} = 0.
\fe
This equation can then be used to solve for $\lambda$ as a function of the temperature $T$ ($x=e^{-\B} = e^{-1/T}$). In the zero temperature limit, we see that $\lambda$ goes to zero. For temperature of order 1, the saddle point value for $\lambda$ is 
\ie
\lambda &= {8\over \pi^2 N} \sum_{m=1}^\infty {x^m(1+x^m)\over (1-x^{m})^3}+ {\cal O}({1\over N^2}).
\fe

At high temperature, $x\sim 1-{1\over T}$ is close to 1. If the temperature is such that $\lambda$ is of order 1, we still have
\ie
z_{S,\lambda}(x^m) \sim z_S(x^m) \sim 2({T\over m})^2,
\fe
and the saddle point equation for $\lambda$ becomes
\ie
\sum_{\ell=0}^\infty \left[1-{\ell+{1\over 2}\over \sqrt{(\ell+{1\over 2})^2+\lambda}}\right]={4T^3\over N}\zeta(3).
\fe
In particular, we see that $\lambda$ is of order 1 when $T$ is of order $N^{1\over 3}$, a rather curious scaling. At this temperature, (\ref{sadt}) takes the form
\ie
\rho(\A) = {1\over 2\pi}+{2T^2 \over \pi N}f(\A),
\fe
where $f(\A)$ is still given by (\ref{falpha}). To see the Gross-Witten transition, we need to go to higher temperature, at which $\lambda$ is large, and the sums over $\ell$ can now be approximated by integrals. (\ref{sadlll}) simplifies to\footnote{Note that this form of the saddle point equation for $\lambda$, derived by assuming (\ref{sadt}), is only valid below the Gross-Witten transition temperature, and in particular is not valid for $T\gg \sqrt{N}$ where the eigenvalues are concentrated near $\A=0$ and one recovers the result of \cite{Sachdev:1993pr}.}
\ie
-\sqrt{\lambda} + {4T^3\over N} \left[ {\sqrt{\lambda}\over T}{\rm Li}_2(e^{-2 \sqrt{\lambda}/T})+{\rm Li}_3(e^{-2 \sqrt{\lambda}/T})\right] = 0.
\fe
So for $T=b\sqrt{N}$, we have $\lambda = \C N$, where $\C$ is determined in terms of $b$ via
\ie\label{gwa}
\sqrt{\C} = 4 b^2 \sqrt{\C}{\rm Li}_2(e^{-2 \sqrt{\C}/b}) + 4 b^3 {\rm Li}_3(e^{-2\sqrt{\C}/b}).
\fe
$z_{S,\lambda}(x^m)$ is now approximated by
\ie
z_{S,\lambda}(x^m) \approx N{2b^2\over m^2} (1+{m\sqrt{\C}\over b}) e^{-m\sqrt{\C}/b},
\fe
and so
\ie
\rho(\A) \approx {1\over 2\pi} + {2b^2\over \pi}\sum_{m=1}^\infty {1+m \sqrt{\C}/b\over m^2}e^{-m\sqrt{\C}/b} \cos(m\A).
\fe
The Gross-Witten transition occurs when $\rho(\A)$ first vanishes at $\A=\pi$, namely
\ie\label{gwb}
{1\over 2\pi} + {2b^2\over \pi}\sum_{m=1}^\infty (-)^m {1+m \sqrt{\C}/b\over m^2}e^{-m\sqrt{\C}/b}  = 0
\fe
(\ref{gwa}) and (\ref{gwb}) can now be solved numerically, giving
\ie
\C \approx 0.140342, ~~~~ b \approx 0.581068.
\fe
To conclude, we find that the Gross-Witten transition of the critical vector model occurs at temperature $T\approx 0.581068 \sqrt{N}$.

In much of the rest of the paper, we will return to the free singlet theory on the sphere.

\section{Relations at finite $N$}
At finite $N$ the singlet bilinears dual to the higher spin bulk states are not independent.  There are nontrivial relations between invariants. As a simple example consider the $O(N)$ theory with $N=1$.  The theory has just one scalar field $\phi(x)$.  Considering products of two bilinears we have the trivial identity
\ie
 (\phi(x) \phi(x)) (\phi(y) \phi(y)) = (\phi(x) \phi(y)) (\phi(x) \phi(y))
 \fe 
 Expanding both sides in $(x-y)$, on the left hand side we have products of descendants of the scalar current $\phi_i \phi_i$, whereas on the right hand side we have products of descendants of general higher spin currents $\phi_i \partial  \ldots \partial  \phi_i$.

We can describe the relations among products of scalar bilinear currents for general $N$ as follows. Consider $\phi_{i_1}(x_1), ..., \phi_{i_m}(x_m)$. These are $mN$ independent objects. If we form $O(N)$ invariants with them, $\phi_i(x_a) \phi_i(x_b)$, we have $m(m+1)/2$ invariants. So when $m>2N-1$, there must be relations among them. This seems to be the minimal $m$ required.

These quantities can be related to the entropy of the system in the following heuristic way.   The characteristic length scale of excitations at temperature $T$ is $1/T$.  Imagine smearing the scalar field operators that create these excitations across  a 2D cell of area $(1/T)^2$.  There are $T^2 V$ such cells on the sphere where $V$ is the 2D volume of the sphere ( not set equal to 1 for now).  Let  $\phi^s_i(x_a)$ be the smeared field operator at the cell entered at $x_a$ with flavor $i$.  These operators create the typical excitations of the system.  There are $N T^2 V$ such operators.  From the above discussion we see we should equate $m \sim  T^2 V$.  At temperature $T$ these operators create states comprising an entropy $S_{hi}= NT^2 V= m N$.   No singlet constraint has been imposed.   This is the entropy of the high temperature phase (\ref{htl}).

Now we examine the $O(N)$ bilinear invariant side, described by smeared invariants of the form $\phi^s_i(x_a) \phi^s_i(x_b)$.  There are $T^4 V^2$ such objects and if they are independent create states comprising an entropy $S_{low} \sim T^4 V^2 = m^2$.  This is the entropy in the low temperature phase\footnote{Really $1 \ll T \ll T_c$.}  (\ref{orderone}).  Note that the somewhat surprising  $T^4 V^2$ dependence is naturally explained here as the independent integral of the two positions of a bilinear operator over the sphere.  

Now the criterion discussed above for relations to occur between invariants is $m N \sim m^2$.    But this occurs for typical states in the thermal ensemble when 
$S_{hi} \sim S_{low}$ or $NT^2 V \sim T^4 V^2$.  This is equivalent, up to order 1 factors, to requiring that the  free energies of the two phases agree.  This determines the phase transition point.  And in fact  this is satisfied when $T \sim \sqrt{N} \sim T_c$ ($V=1$ here).   So the large $N$ transition is the place at which relations become significant between the bilinear invariants.  At large $T$ relations dominate so the true entropy $S_{hi}$ is much less than the independent invariant count $S_{low}$.   This analysis applies in arbitrary dimensions.

We can now estimate the lowest scaling dimension $\Delta$ operator in the field theory that obeys nontrivial relations. 
Such a relation involves the product of at least $ 2m$ fields, with $m $ different positions; we can then expand the relation in powers of $ x_{21}^\mu, x_{31}^\mu, ..., x_{m1}^\mu$, where $x_{ij}^\mu=x_i^\mu-x_j^\mu$. In other words, we may express the operator relation in the form
$F(x_{21}, x_{31}, ..., x_{m1}) = 0$,
where the coefficients of the polynomial in $x_{i1}$ are operators made out of the bilinear currents at $x_1$.

But for the minimal value $m=2N$, this function $F$ must have the property that when a pair $x_i$ and $x_j$ are set equal, the relation becomes trivial. This essentially implies that $F$ must have at least the same degree as a completely antisymmetric polynomial function $G$. The counting of degree of $F$ becomes the same as the counting of the dimension of the baryon operator discussed below, which scales like $N^{3/2}$.  So we expect the lowest $\Delta$ obeying nontrivial relations to be of order $N^{3/2}$.

In fact, by directly inspecting the difference between partition function $Z_{U(N)}(x)$ of the $U(N)$-invariant vector model ($x=e^{-1/T}$) and that of the free higher spin gas in $AdS_4$, $Z_{HS}(x)$ (given by exponentiating (\ref{freeHSgas})), up to $N=8$, we find the following formula
\ie
Z_{HS}(x) - Z_{U(N)}(x) = {2n+1\choose r}^2 \,x^{{4n^3-n\over 3}+(2n+1)r} + {\rm higher ~order~in}~x,~~~N=n^2+r,~~n=\lfloor \sqrt{N}\rfloor.
\fe
We conjecture that this holds for general $N$. This means that nontrivial relations among the bilinear invariants first appear at level ${4n^3-n\over 3}+(2n+1)r$ ($\sim {4\over 3}N^{3\over 2}$ at large $N$), and that there are ${2n+1\choose r}^2$ relations at this level.

In the $SU(N)$ and $SO(N)$ theory there  are invariant operators not related to bilinears built using the $\epsilon$ tensor.
These involve order $N$ boson or fermion fields.  For fermions a natural candidate is 
\ie
{\cal O}_f = \epsilon^{i_1 \ldots i_N}\epsilon^{j_1\ldots j_N} \psi^{\A_1}_{i_1}\ldots\psi^{\A_N}_{i_N}\psi_{j_1 \A_1} \ldots \psi_{\j_N \A_N}
\fe
This operator has $\Delta = 2N$.
 
For bosons a natural operator would have the schematic form
\ie
{\cal O}_b = \epsilon^{i_1 \ldots i_N} \phi_{i_1} \partial \phi_{i_2} ....\partial \partial \ldots \phi_{i_N}
\fe
Where the derivatives are arranged to give a nonvanishing contribution.  There are $d$ derivatives $\partial_\mu$ that can be distributed on the $N$ $\phi_i$'s, which are symmetric on the same site and antisymmetric between different sites. We also need to take into account the (free) equation of motion on each $\phi_i$.
 
 Writing the single scalar letter partition function as
 \ie
 {1+x\over (1-x)^{d-1}} = \sum a_n x^n,
 \fe
 the partition function for the operators of type ${\cal O}_b$ is
 \ie
 Z_{d,N} (x) & = x^{N\over 2}\left. \prod_{n=0}^\infty (1+x^n y)^{a_n}\right|_{y^N} 
 = \left.e^{\sum_{m=1}^\infty {(-)^{m-1}\over m} y^m z_S(x^m)}\right|_{y^N}
\fe
In the $d=3$ case, we have $a_n=2n+1$, and so
\ie
Z_{3,N}(x) = x^{N\over 2} \left. \prod_{n=0}^\infty (1+x^n y)^{2n+1}\right|_{y^N}
\fe
The term at a given order in $y$, and of lowest order in $x$, is obtained from
\ie
x^{N\over 2} \prod_{n=0}^{k-1} (x^n y)^{2n+1} = x^{{N\over 2}+{k(k-1)(4k+1)\over 6}} y^{k^2}
\fe
Identifying $k^2=N$, the exponent of $x$ goes like
\ie
\Delta \sim {2\over 3} N^{3\over 2}.
\fe
In general $d$ dimensions, we have
\ie
\sum_{n=0}^{k-1} a_n \sim {2k^{d-1}\over (d-1)!}\sim N,~~~~ \Delta \sim \sum_{n=0}^{k-1} n a_n
\sim {d-1\over d} k N,
\fe
and so the dimension of the operator ${\cal O}_b$ scales like
\ie
\Delta \sim  N^{d\over d-1}.
\fe
or $N^{3/2}$ in $d=3$.  
At temperatures of order 1  we expect leading corrections to the higher spin gas thermal free energy   to be of order $\exp(- \Delta/ T) $, nonperturbatively small in $N$.  This gives $\exp(-N^{3/2}/T)$ for the scalar theory and $\exp(-N/T)$ for fermions.
 
The above arguments were made directly in the field theory state space.   There also exist some arguments about nonperturbative effects in $N$ in the matrix model.
 In the simplest matrix model with $\cos (\A)$ potential  in the analog of the low temperature phase the corrections to the leading answer are $\exp(-N)$ \cite{Goldschmidt:1979hq}. 
 
  For $b < b_c$  standard unitary matrix models  in the double scaled limit \cite{Periwal:1990gf}  are known to have leading exponentially small corrections \cite{Crnkovic:1990ms, Watterstam:1990qs} to the free energy. 
  For $b > b_c$  standard models have perturbative $1/N$ effects and  $\exp(-N)$ effects due to one eigenvalue instantons \cite{Neuberger:1980qh, David:1990sk, Shenker:1990}.   We expect these here. The gravitational interpretation of these effects needs to be understood.

\section{Discussion}

We have seen that the higher spin AdS gas thermodynamics describes the system well for all temperatures up to $T_c \sim \sqrt{N}$, the Planck scale.     At  $T=T_c$ relations between the bilinear invariants describing the higher spin states become important and so above $T_c$  the number of degrees of freedom is less than in the higher spin gas.\footnote{This reduction of degrees of freedom is reminiscent of the $AdS_3$ higher spin gas studied in \cite{Castro:2011ui}.}

The bulk interpretation of these results is unclear.  The large AdS black hole familar from ordinary AdS/CFT must either be absent in this theory  or have subdominant free energy for lower temperatures.  
The known black hole solution of Didenko-Vasiliev \cite{Didenko:2009td} is an extremal charged black hole, whose charges are not yet understood.  It is not a candidate for the generic thermal state.  

If the large AdS black holes had subdominant free energy in the $2+1$ D boundary theory  we would expect them to give nonperturbative effects of strength $\exp(-N/T)$ since $G_N \sim 1/N$.     But as we argued in section 5 the leading effects in the scalar theory are of order  $\exp(- N^{3/2}/T)$, too small to be due to black holes.  The $U(N)$ fermion theory has effects of order $\exp(-N/T)$.  Their bulk origin is unclear.   The $SU(N)$ fermion theory has  a charged state with $\Delta = 2N$.  This could be a BPS object.   Preliminary investigations suggest that the number of excited states that can be created above it are not enough to describe a nonextremal black object.

Following the conjecture \cite{Gaberdiel:2010pz}  relating a  $1+1D$ $W_N$ minimal model to a $2+1$ dimensional higher spin gauge theory (coupled to scalar matter), the authors of \cite{Kraus:2011ds} recently connected bulk solutions resembling BTZ black holes to boundary field theory entropy.  The high temperature states at $\lambda = 1$ where $\lambda$ is the `t Hooft coupling $N/k$ correspond to the singlet projected scalar.\footnote{We ignore complications from the zero mode, or alternatively we can consider the $\lambda =0$ free fermion theory.} But there are known to be extra light  states in this system that do not decouple \cite{Papadodimas:2011pf} and will affect finite temperature thermodynamics and possible large $N$ phase transitions.  In fact the authors of  \cite{Papadodimas:2011pf} find an entropy proportional to $N$ at arbitrarily low $T$.\footnote{For all $\lambda \neq 0$.}  A simple way to see the necessity of more degrees of freedom is to remember that the $W_N$ CFT is modular invariant. But the partition function with a holonomy on a timelike cycle projecting on singlet states is not modular invariant.  A modular transform transfers the holonomy onto a spacelike cycle, altering the spatial boundary conditions and producing new light  states in ``twisted" sectors with a continuous twist.\footnote{We thank Mathias Gaberdiel for explaining this to us.}

  The local Lagrangian description of the $2+1D$ boundary theory using a Chern-Simons gauge field is useful in defining the theory on general manifolds, particularly ones of nontrivial topology. Alternatively, one may directly enforce the singlet constraint on a general 3-manifold by coupling the scalar fields to a flat connection, and integrate over all flat connections.   The bulk interpretation of these flat connection degrees of freedom is an open, interesting question, possibly related to the twisted sectors in $1+1$ dimensions discussed above.

\section*{Acknowledgements}

We would like to thank Chi-Ming Chang, Vyatcheslav Didenko, Mathias Gaberdiel, Simone Giombi, Daniel Jafferis, Per Kraus, Massimo Porrati, Andy Strominger and Mikhail Vasiliev for helpful discussions and correspondences.   The research of SS is supported by NSF Grant 0756174 and the Stanford Institute for Theoretical Physics.  The research of XY is supported by the Fundamental Laws Initiative Fund at Harvard University, and in part by NSF Award PHY-0847457.   We would like to thank the Aspen Center for Physics where this work was initiated.

\end{document}